\documentclass[preprint,12pt]{elsarticle}
\usepackage{graphicx}
\usepackage{amsmath}
\usepackage{amssymb}

\newcommand{\vecr}[1]{\mathbf{#1}}

\newcommand{\vecrsym}[1]{\boldsymbol{#1}}

\newcommand{\T}[1]{{#1}^{\text{T}} }

\newcommand{\norm}[2]{ \left| \left| #1 \right| \right|_{#2} }

\newcommand{\dpart}[2]{\frac{\partial #1}{\partial #2}}

\usepackage{xcolor}

\newcommand{\blue}[1]{\textcolor{blue}{#1}}

\usepackage{subfig}
\usepackage{ mathrsfs } 
\DeclareMathAlphabet{\mathscrbf}{OMS}{mdugm}{b}{n}

\journal{arXiv.org}

\begin{document}

\begin{frontmatter}

\title{Large scale three-dimensional topology optimisation of heat sinks cooled by natural convection}

\author[auth1]{Joe Alexandersen\corref{cor1}}
\ead{joealex@mek.dtu.dk}
\cortext[cor1]{Corresponding author}
\author[auth1]{Ole Sigmund}
\author[auth1]{Niels Aage}
\address[auth1]{Department of Mechanical Engineering, Solid Mechanics \\ Technical University of Denmark \\ Nils Koppels All\'e, Building 404 \\ DK-2800, Denmark}

\begin{abstract}
This work presents the application of density-based topology optimisation to the design of three-dimensional heat sinks cooled by natural convection. The governing equations are the steady-state incompressible Navier-Stokes equations coupled to the thermal convection-diffusion equation through the Bousinessq approximation.
The fully coupled non-linear multiphysics system is solved using stabilised trilinear equal-order finite elements in a parallel framework allowing for the optimisation of large scale problems with order of 40-330 million state degrees of freedom.
The flow is assumed to be laminar and several optimised designs are presented for Grashof numbers between $10^3$ and $10^6$. Interestingly, it is observed that the number of branches in the optimised design increases with increasing Grashof numbers, which is opposite to two-dimensional optimised designs.

\end{abstract}


\end{frontmatter}

\section{Introduction} \label{sec:intro}

Natural convection is the phenomena where density-gradients due to temperature differences cause a
fluid to move. Natural convection is therefore a natural way to passively cool a hot object, such as
electronic components, light-emitting diode lamps or materials in food processing.

Topology optimisation is a material distribution method \cite{Bendsoee2003} used to optimise the
layout of a structure in order to minimise a given performance measure subject to design
constraints and a physical model. In order to take convective heat transfer, to an ambient fluid, into account in the design process of density-based methods, a common extension is to introduce some form of interpolation of the convection boundaries, see e.g. \cite{Yin2002,Bruns2007,Iga2009}. More recently, Dede et al. \cite{Dede2015} used these simplified models to design and manufacture heat sinks subject to jet impingement cooling.
However, it is hard to justify the application of a predetermined and constant convection coefficient, because topology optimisation often leads to unanticipated designs, closed cavities and locally varying velocities. During the optimisation process, the design also changes significantly and the interaction with the ambient fluid changes. Therefore, to ensure physically correct capturing of the aspects of convective heat transfer, the full conjugate heat transfer problem must be solved.

Topology optimisation for fluid systems began with the treatment of Stokes flow in the seminal article by Borrvall and Petersson \cite{Borrvall2003} and has since been applied to Navier-Stokes \cite{Gersborg-Hansen2005}, as well as scalar transport problems \cite{Andreasen2009} amongst others. The authors have previously presented a density-based topology optimisation approach for two-dimensional natural convection problems \cite{Alexandersen2014}. Recently, Coffin and Maute presented a level-set method for steady-state and transient natural convection problems using X-FEM \cite{Coffin2015}. Interested readers are referred to \cite{Alexandersen2014} for further references on topology optimisation in fluid dynamics and heat transfer.

Throughout this article, the flows are assumed to be steady and laminar. The fluid is assumed to be incompressible, but buoyancy effects are taken into account through the Boussinesq approximation, which introduces variations in the fluid density due to temperature gradients. The inclusion of a Brinkman friction term facilitates the topology optimisation of the fluid flow.

The scope of this article is primarily to present large scale three-dimensional results using the formulation presented in \cite{Alexandersen2014}, as well as the computational issues arising from solving the non-linear equation system and the subsequent linear systems. Thus, only a brief overview of the underlying finite element and topology optimisation formulations are given and the reader is referred to \cite{Alexandersen2014} for further information.

In recent years an increasing body of work has been published on efficient large scale topology optimisation. These works cover the use of high-level scripting languages \citep{Andreassen2011,Amir2013}, multiscale/-resolution approaches \citep{Alexandersen2015,Lee2011c} and parallel programming using the message parsing interface and C/Fortran \cite{Borrvall2001,Evgrafov2008,Aage2013,Aage2014}.
To facilitate the solution to truly large scale conjugate heat transfer problems, the implementation in this paper is done using PETSc \cite{petsc2014} and the framework for topology optimisation presented in \cite{Aage2014}.

The layout of the paper is as follows: Section \ref{sec:governingEquations} presents the governing equations; Section \ref{sec:optimisation} presents the topology optimisation problem; Section \ref{sec:formulation} briefly discusses the finite element formulation; Section \ref{sec:numerical} discusses the numerical implementation details; Section \ref{sec:results} presents scalability results for the parallel framework and optimised designs for a test problem; Section \ref{sec:conclusions} finishes with a discussion and conclusion.

\section{Governing equations} \label{sec:governingEquations}

The dimensionless form of the governing equations have been derived based on the Navier-Stokes and convection-diffusion equations under the assumption of constant fluid properties, incompressible, steady flow and neglecting viscous dissipation. Furthermore, the Boussinesq approximation has been introduced to take density-variations due to temperature-differences into account.
A domain is decomposed into two subdomains, $\Omega = \Omega_{f} \cup \Omega_{s} $, where $\Omega_{f}$ is the fluid domain and $\Omega_{s}$ is the solid domain.
In order to facilitate the topology optimisation of conjugate natural convective heat transfer between a solid and a surrounding fluid, the equations are posed in the unified domain, $\Omega$, and the subdomain behaviour is achieved through the control of coefficients. The following dimensionless composite equations are the result.
\\ $\forall \vecr{x} \in \Omega:$
\begin{align}
u_{j} \frac{\partial u_{i}}{\partial x_{j}} - {Pr}\frac{\partial}{\partial x_{j}} \left( \frac{\partial u_{i}}{\partial x_{j}} + \frac{\partial u_{j}}{\partial x_{i}} \right) + \frac{\partial p}{\partial x_{i}} &= -\alpha(\vecr{x}) u_{i} - {Gr}{Pr}^{2}\,e^{g}_{i}\,T \label{eq:BoussDimless-a}\\
\frac{\partial u_{j}}{\partial x_{j}} &= 0 \label{eq:BoussDimless-b}\\
u_{j} \frac{\partial T}{\partial x_{j}} - \frac{\partial
}{\partial x_{j}} \left( K(\mathbf{x}) \frac{\partial T}{\partial x_{j}} \right) &= s(\vecr{x}) \label{eq:BoussDimless-c}
\end{align}
where $u_{i}$ is the velocity field, $p$ is the pressure field, $T$ is the temperature field, $x_{i}$ denotes the spatial coordinates, $e^{g}_{i}$ is the unit vector in the gravitational direction, $\alpha(\vecr{x})$ is the spatially-varying effective impermeability, $K(\vecr{x})$ is the spatially-varying effective thermal conductivity, $s(\vecr{x})$ is the spatially-varying volumetric heat source term, $Pr$ is the Prandtl number, and $Gr$ is the Grashof number.

The effective thermal conductivity, $K(\vecr{x})$, is defined as:
\begin{equation} \label{eq:conducends}
K(\vecr{x}) = \left\lbrace
\begin{matrix}
1 & \text{ if }\, \vecr{x} \in \Omega_{f} \\
\frac{1}{C_{k}} & \text{ if }\, \vecr{x} \in \Omega_{s}
\end{matrix} \right.
\end{equation}
where $C_{k} = \frac{k_{f}}{k_{s}}$ is the ratio between the fluid thermal conductivity, $k_{f}$, and the solid thermal conductivity, $k_{s}$. Theoretically, the effective impermeability, $\alpha(\vecr{x})$, is defined as:
\begin{equation} \label{eq:alphaends}
\alpha(\vecr{x}) = \left\lbrace
\begin{matrix}
0 & \text{ if }\, \vecr{x} \in \Omega_{f} \\
\infty & \text{ if }\, \vecr{x} \in \Omega_{s}
\end{matrix} \right.
\end{equation}
in order to ensure zero velocities inside the solid domain. However, numerically this requirement must be relaxed as will be described in section \ref{sec:optimisation}.
The volumetric heat source term is defined as being active within a predefined subdomain of the solid domain, $\omega \subset \Omega_{s}$:
\begin{equation}
s(\vecr{x}) = \left\lbrace
\begin{matrix}
0 & \text{ if }\, \vecr{x} \notin \omega \\
s_{0} & \text{ if }\, \vecr{x} \in \omega
\end{matrix} \right.
\end{equation}
where $s_{0}=\frac{qL}{k_{s}\Delta T}$ is the dimensionless volumetric heat generation, $q$ is the dimensional volumetric heat generation, $\Delta T$ is the reference temperature difference and $L$ is the reference length scale.

The Prandtl number is defined as:
\begin{equation}
Pr = \frac{\nu}{\Gamma}
\end{equation}
where $\nu$ is the kinematic viscosity, or momentum diffusivity, and $\Gamma$ is the thermal diffusivity. It thus describes the relative spreading of viscous and thermal effects.
The Grashof number is defined as:
\begin{equation}
Gr = \frac{g \beta \,{\Delta T}\,L^{3}}{\nu^{2}}
\end{equation}
where $g$ is the acceleration due to gravity and $\beta$ is the volumetric coefficient of thermal expansion. It describes the ratio between the buoyancy and viscous forces in the fluid. The Grashof number is therefore used to describe to what extent the flow is dominated by natural convection or diffusion. For low $Gr$ the flow is dominated by viscous diffusion and for high $Gr$ the flow is dominated by natural convection.
The problems in this article are assumed to have large enough buoyancy present to exhibit natural convective effects, but small enough $Gr$ numbers to exhibit laminar fluid motion.

\section{Optimisation formulation} \label{sec:optimisation}

\subsection{Interpolation functions}
In order to perform topology optimisation, a continuous design field, $\gamma(\vecr{x})$, varying between 0 and 1 is introduced. Pure fluid is represented by $\gamma(\vecr{x}) = 1$ and solid by $\gamma(\vecr{x}) = 0$. For intermediate values between 0 and 1, the effective conductivity is interpolated as follows:
\begin{equation} \label{eq:conducint}
K( \gamma ) = \frac{\gamma ( C_{k} ( 1 + q_{f} ) -1 ) + 1}{ C_{k} ( 1 + q_{f} \gamma ) }
\end{equation}
and likewise the effective impermeability is interpolated using:
\begin{equation} \label{eq:alphaint}
\alpha\left( \gamma \right) = \overline{\alpha} \frac{1-\gamma}{1+q_{\alpha}\gamma}
\end{equation}
The interpolation functions ensure that the end points defined in \eqref{eq:conducends} and \eqref{eq:alphaends}, respectively, are satisfied.
The effective impermeability is bounded to $\overline{\alpha}$ in the solid regions and this upper bound should be chosen large enough to provide vanishing velocities, but small enough to ensure numerical stability.
The convexity factors, $q_{f}$ and $q_{\alpha}$, are used to control the material properties for intermediate design values in order to promote well-defined designs without intermediate design field values.

\subsection{Optimisation problem}
The topology optimisation problem is defined as:
\begin{align} \label{eq:topopt_prob}
\underset{ \vecrsym{\gamma} \in \mathcal{D} }{\text{minimise:}} & f\negthinspace \left( {\gamma}, T \right) = \int_{\omega}\negmedspace{ s(\vecr{x})\,T }\,dV \nonumber\\[-0.1cm]
\text{subject to: } & g\negthinspace \left( {{\gamma}}\right) = \int_{\Omega_{d}}\negmedspace{ 1 - \gamma }\,dV \leq v_{f}\negthinspace\int_{\Omega_{d}}{  }dV \\
 & \mathscr{R} \! \left( {{\gamma}}, \vecr{u}, p, T \right) = {0} \nonumber \\
 & 0 \leq \gamma(\vecr{x}) \leq 1 \,\,\,\, \forall \vecr{x}\in \Omega_{d} \nonumber
\end{align}
where ${\gamma}$ is the design variable field, $\mathcal{D}$ is the design space, $f$ is the thermal compliance objective functional, $g$ is the volume constraint functional,
${\mathscr{R}}\! \left( {{\gamma}}, \vecr{u}, p, T \right)$ is the residual arising from the stabilised weak form of the governing equations, and $\Omega_{d}\subseteq\Omega$ is the design domain.

The design field is regularised using a PDE-based (partial differential equation) density filter \cite{Lazarov2011,Aage2014} and the optimisation problem is solved using the method of moving asymptotes (MMA) \cite{Svanberg1987,Aage2014}.

\subsection{Continuation scheme}
A continuation scheme is performed on various parameters in order to stabilise the optimisation process and to improve the optimisation results. It is the experience of the authors that the provided continuation scheme yields better results than starting with the end values, although this cannot generally be proven \cite{Stolpe2001,Rojas-Labanda2015}.

The chosen continuation strategy consists of five steps:
\begin{subequations}
\begin{align}
q_{f} \in &\left\lbrace 0.881, 8.81, 88.1, 881, 881 \right\rbrace \\
q_{\alpha} \in &\left\lbrace 8, 8, 8, 98, 998 \right\rbrace \\
\overline{\alpha} \in &\left\lbrace 10^{5}, 10^{5}, 10^{5}, 10^{6}, 10^{7} \right\rbrace
\end{align}
\end{subequations}
The sequence is chosen in order to alleviate premature convergence to poor local minimum. The value of $q_{f}$ is slowly increased to penalise intermediate design field values with respect to conductivity. The maximum effective permeability, $\overline{\alpha}$, is set relatively low during the first three steps, as this ensures better scaled sensitivities and more stable behaviour. Over the last two steps, $\overline{\alpha}$ is increased by two orders of magnitude in order to further decrease the velocity magnitudes in the solid regions. The particular values of $q_{f}$ are chosen by empirical inspection such as to ensure the approximate collocation of the fluid and thermal boundaries.
A more comprehensive investigation into the modelling accuracy of the boundary layers is outside the scope of this paper and will be investigated in future work.

It is important to note that the optimisation problem is by no means convex and any optimised design will at best be a local minimum. The obtained design will always depend on the initial design as well as the continuation strategy. However, in the authors experience, the chosen continuation strategy gives a good balance between convergence speed, final design performance and physicality of the modelling. The effects of the steps of the current continuation strategy on the design distribution will be discussed in section \ref{sec:results_hugeDes}.

\section{Finite element formulation} \label{sec:formulation}

The governing equations are discretised using stabilised trilinear hexahedral finite elements. The design field is discretised using elementwise constant variables, in turn rendering the effective thermal conductivity and the effective impermeability to be elementwise constant.
The monolithic finite element discretisation of the problem ensures continuity of the temperature field, as well as the fluxes across fluid-solid interfaces.
The particularities of the implemented finite element formulation are as detailed in \cite{Alexandersen2014}. However, simpler stabilisation parameters have been used in order to allow for consistent sensitivities to be employed, see \ref{app:stabilisation}. The Jacobian matrix is now fully consistent in that variations of the stabilisation parameters with respect to design and state fields are included, in contrast to the original work \citep{Alexandersen2014}.

\section{Numerical implementation} \label{sec:numerical}

The discretised FEM equation is implemented in PETSc \citep{petsc2014} based on the topology optimisation framework presented in \citep{Aage2014}. The PETSc framework is used due to its parallel scalability, the availability of both linear and non-linear solvers, preconditioners and structured mesh handling possibilities. All components described in the following are readily available within the PETSc framework.

\subsection{Solving the non-linear system} \label{sec:solvingNonlinear}

The non-linear system of equations is solved using a damped Newton method. The damping coefficient is updated using a polynomial $L^2$-norm fit, where the coefficient is chosen as the minimiser of the polynomial fit. The polynomial fit is built using the $L^2$-norm of the residual vector at the current solution point, at 50\% of the Newton step and at 100\% of the Newton step. This residual-based update scheme combined with a good initial vector (the solution from the previous design iteration) has been observed to be very robust throughout the optimisation process for the moderately non-linear problems treated. To further increase the robustness of the non-linear solver, if the Newton solver fails from the supplied initial vector, a ramping scheme for the heat generation magnitude is applied in order to recover. Throughout, the convergence criteria for the Newton solver is set to a reduction in the $L^2$-norm of the residual of $10^{-4}$ relative to the initial residual.

\subsection{Solving the linear systems} \label{sec:solvingLinear}

Due to the large scale and three-dimensional nature of the treated problems, the arising linearised systems of equations are by far the most time consuming part of the Newton scheme. Therefore, to make large scale problems tractable the (unsymmetric) linear systems are solved using a fully parallelised iterative Krylov subspace solver.

Constructing an iterative solver that is both independent of problem settings and possesses both parallel and numerical scalability, is intricate and beyond the scope of this work, where focus is on the optimisation. However, in order to facilitate the solution of large scale optimisation problems, an efficient solver is required. To this end the authors use the readily available core components in PETSc to construct a solver with focus on simplicity, as well as reduced  wall clock time. This is quite easy to obtain for solvers and preconditioners that rely heavily on matrix-vector multiplications. To this purpose the flexible GMRES (F-GMRES) method is used as the linear solver combined with a geometry-based Galerkin-projection multigrid (GMG) preconditioner. Such a solver depends highly on the quality of the smoother to guarantee fast convergence \citep{Vassilevski2007}. The authors have found that a simple Jacobi-preconditioned GMRES provides a reasonable choice of smoother\footnote{Due to the choice of a Krylov smoother, the multigrid preconditioner will vary slightly with input and thus require a flexible outer Krylov method.}. The convergence criterion for the Krylov solver is set to $10^{-5}$ relative to the initial residual.

\section{Results} \label{sec:results}

\subsection{Problem setup} \label{sec:results_cavity}

\begin{figure}
\centering
\subfloat[Dimensions]{\includegraphics[width=0.25\textwidth]{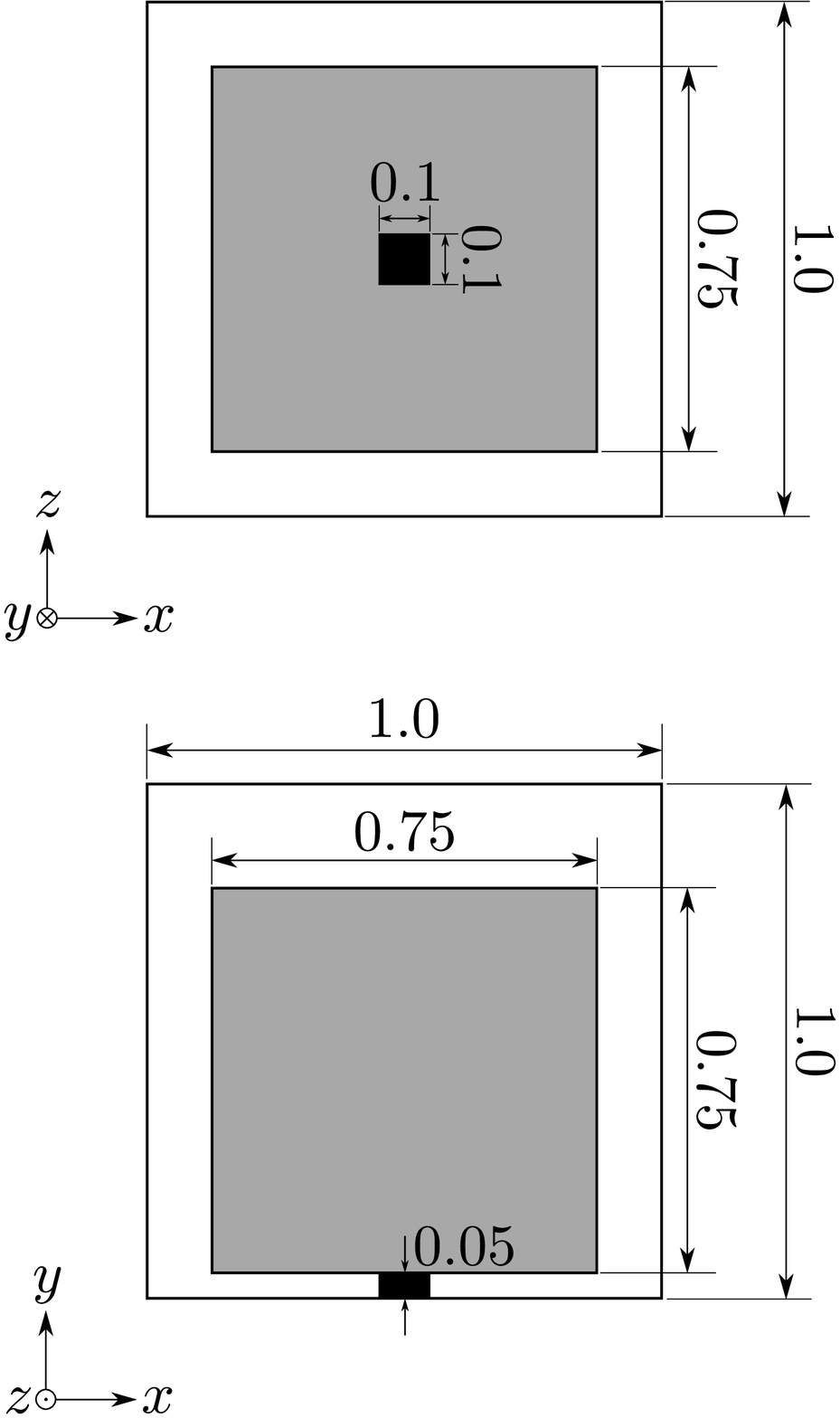}
\label{fig:cavity_probSetup-a}}
\subfloat[Boundary conditions]{\includegraphics[width=0.25\textwidth]{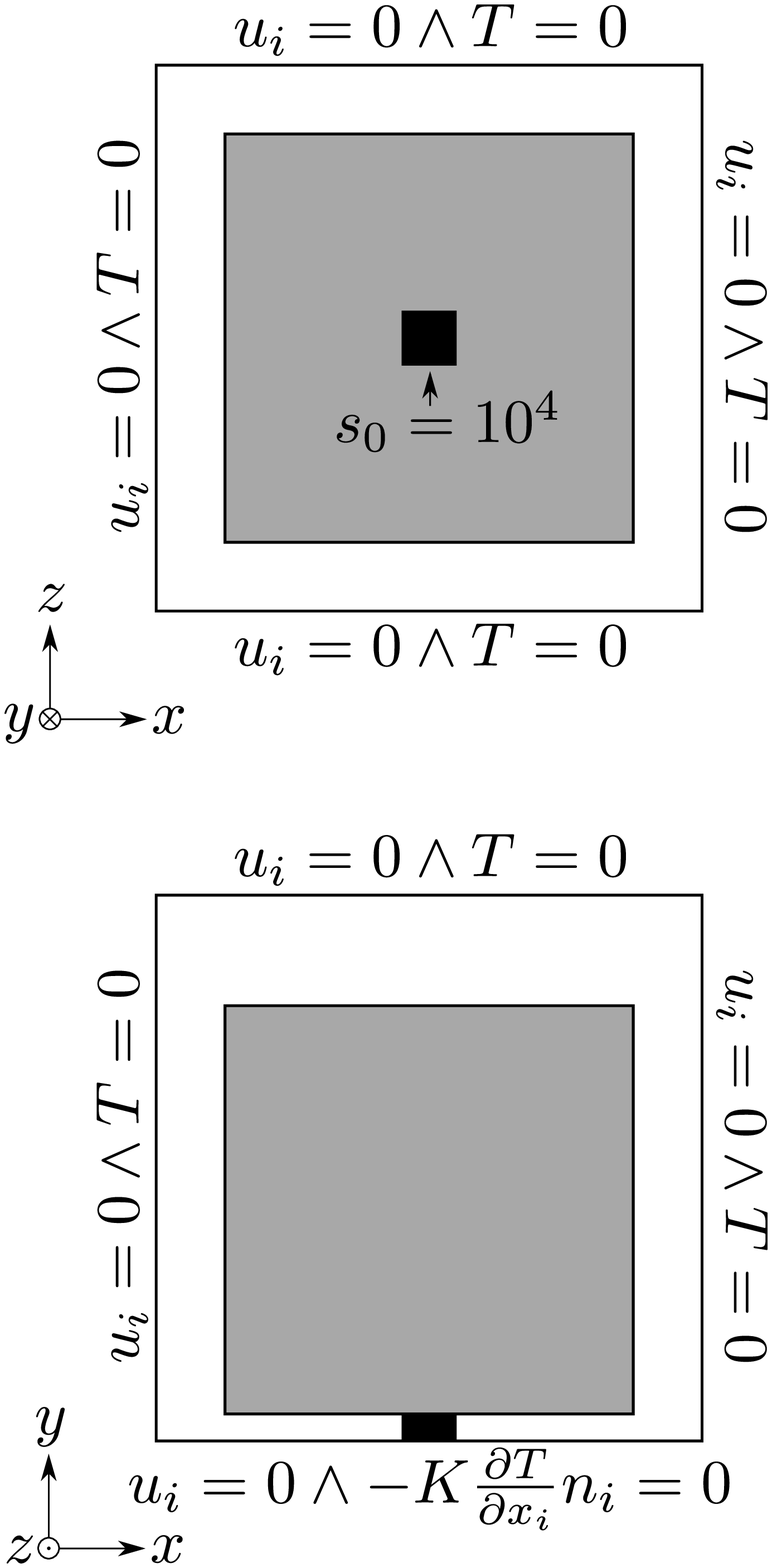}
\label{fig:cavity_probSetup-b}}
\caption{Illustration of the problem setup for the heat sink in a close cubic cavity. The heat source is black and the design domain is gray.}\label{fig:cavity_probSetup}
\end{figure}

The considered optimisation problem is an (academic) example of a heat sink in a closed cubic cavity. Figure \ref{fig:cavity_probSetup} shows the problem setup with dimensions and boundary conditions. The heat source (black in figure), exemplifying an electronics chip, is placed in the mid-bottom of the cavity and modelled using a small block of solid material with volumetric heat generation, $s_{0}=10^{4}$. The design domain (gray in figure) is placed on top of the heat source in order to allow the cooling fluid to pass underneath it, as well as to allow room for wires etc. The vertical and top outer walls of the cavity are assumed to be kept at a constant cold temperature, $T=0$, while the bottom is insulated.  The height of the cavity has been used as the reference length scale, $L$. Thus, the cavity dimensions are $1\times1\times1$, the design domain dimensions are $0.75\times0.75\times0.75$ and the heat source dimensions are $0.1\times0.05\times0.1$. A discussion on the definition of the temperature difference in the Grashof number can be found in \ref{app:grashofNumber}.
Initial investigations showed that due to the symmetry of the domain and boundary conditions, the design and state solutions remained quarter symmetric throughout the optimisation. Thus, the computational domain is limited to a quarter of the domain with symmetry boundary conditions. The volume fraction is set to 5\%, i.e. $v_{f}=0.05$, for all examples.

\subsection{Parallel performance}
To demonstrate the parallel performance of the state solver, the model optimisation problem described in section \ref{sec:results} is solved on a fixed mesh for different numbers of processes. All computations in this paper were performed on a cluster, where each node is equipped with two Intel Xeon e5-2680v2 10-core 2.8GHz processors. The results shown in Tables \ref{tab:scalabilityGr1e3} and \ref{tab:scalabilityGr1e6} are averaged over 250 design cycles and show the performance for $Gr=10^{3}$ and $Gr=10^{6}$, respectively.
The data shows that the proposed solver scales almost linearly in terms of speed up, and more importantly that the performance is only slightly affected by the Grashof number.
\begin{table}
\centering
\begin{tabular}{ccc}
Processes &  time [s] &  scaling \\ \hline
160 &  53.2 & 1.00 \\
320 &  28.9 & 0.54 \\
640 &  14.1 & 0.26 \\
\end{tabular}
\caption{Average time taken per state solve over 250 design iterations for $Gr=10^{3}$ at a mesh resolution of $80\times160\times80$.} \label{tab:scalabilityGr1e3}
\end{table}
\begin{table}
\centering
\begin{tabular}{cccc}
Processes & time [s] &  scaling \\ \hline
160 &  62.6 & 1.00 \\
320 &  31.9 & 0.51 \\
640 &  16.5 & 0.26 \\
\end{tabular}
\caption{Average time taken per state solve over 250 design iterations for $Gr=10^{6}$ at a mesh resolution of $80\times160\times80$.} \label{tab:scalabilityGr1e6}
\end{table}

\begin{table}
\centering
\begin{tabular}{ccc}
Mesh size &  $Gr = 10^{3}$ &  $Gr = 10^{6}$ \\ \hline
$80 \times 160 \times 80$ & 7.5 & 5.6 \\
$160 \times 320 \times 160$ & 10.1 & 7.7 \\
$320 \times 640 \times 320$ & 18.4 & 15.6 \\
\end{tabular}
\caption{Average iterations for the linear solver per state solve over entire design process for $Gr=10^{3}$ and $Gr=10^{6}$ at varying mesh resolutions.} \label{tab:numscal}
\end{table}
In order to quantify the degree of numerical scalability, a second study is performed in which the mesh resolution is varied. The study is conducted for both low and high Grashof numbers and the average F-GMRES iterations are collected in Table \ref{tab:numscal}. The total number of design iterations averaged over was 250, 500 and 1000, respectively, for the three mesh resolutions.
The data clearly shows that the computational complexity increases with problem size, and thus that the solver is not numerically scalable. However, since the growth in numerical effort, i.e. the number of F-GMRES iterations, is moderate we conclude that the proposed solver is indeed applicable for solving large scale natural convection problems.

\subsection{Varying Grashof number} \label{sec:results_varGr}

The problem is investigated for varying $Gr$ under constant volumetric heat generation, $s_{0}=10^{4}$, Prandtl number, $Pr=1$, and thermal conductivity ratio, $C_{k}=10^{-2}$. The purpose of the present study is not to provide a detailed physical example. It is rather to provide phenomenological insight into the effect of changing the governing parameter of the fluid-thermal coupling, namely the Grashof number. However, physical interpretations of tuning the $Gr$-number, while keeping the dimensionless volumetric heat generation and the $Pr$-number constant, could be e.g. equivalent to tuning the gravitational strength (going from microgravity towards full gravity) or the dimensional volumetric heat generation. It is important to note that when interpreting the results, the dimensional temperature scale will differ for the two interpretations. While tuning the gravitational strength, the temperature scale remains the same; by varying the magnitude of the volumetric heat generation, the temperature scale varies accordingly\footnote{The dimensionless volumetric heat generation is given by $s_{0}=\frac{qL}{k_{s}\Delta T}$. Requiring that $s_{0}$ remains constant, means that the dimensional volumetric heat generation, $q$, and the reference temperature difference, $\Delta T$, must vary in unison, i.e. $\frac{q}{\Delta T}=\text{const.}$. An increase in the $Gr$-number is thus achieved through the increase of $\Delta T$.}.

The computational mesh is $160\times320\times160$ elements yielding a total of $8,192,000$ elements and $41,603,205$ degrees of freedom for the quarter domain. The design domain consists of $3,456,000$ elements and the filter radius is set to $2.5$ times the element size.

\begin{figure*}
\centering
\subfloat[$Gr=10^{3}$]{\includegraphics[width=0.45\textwidth]{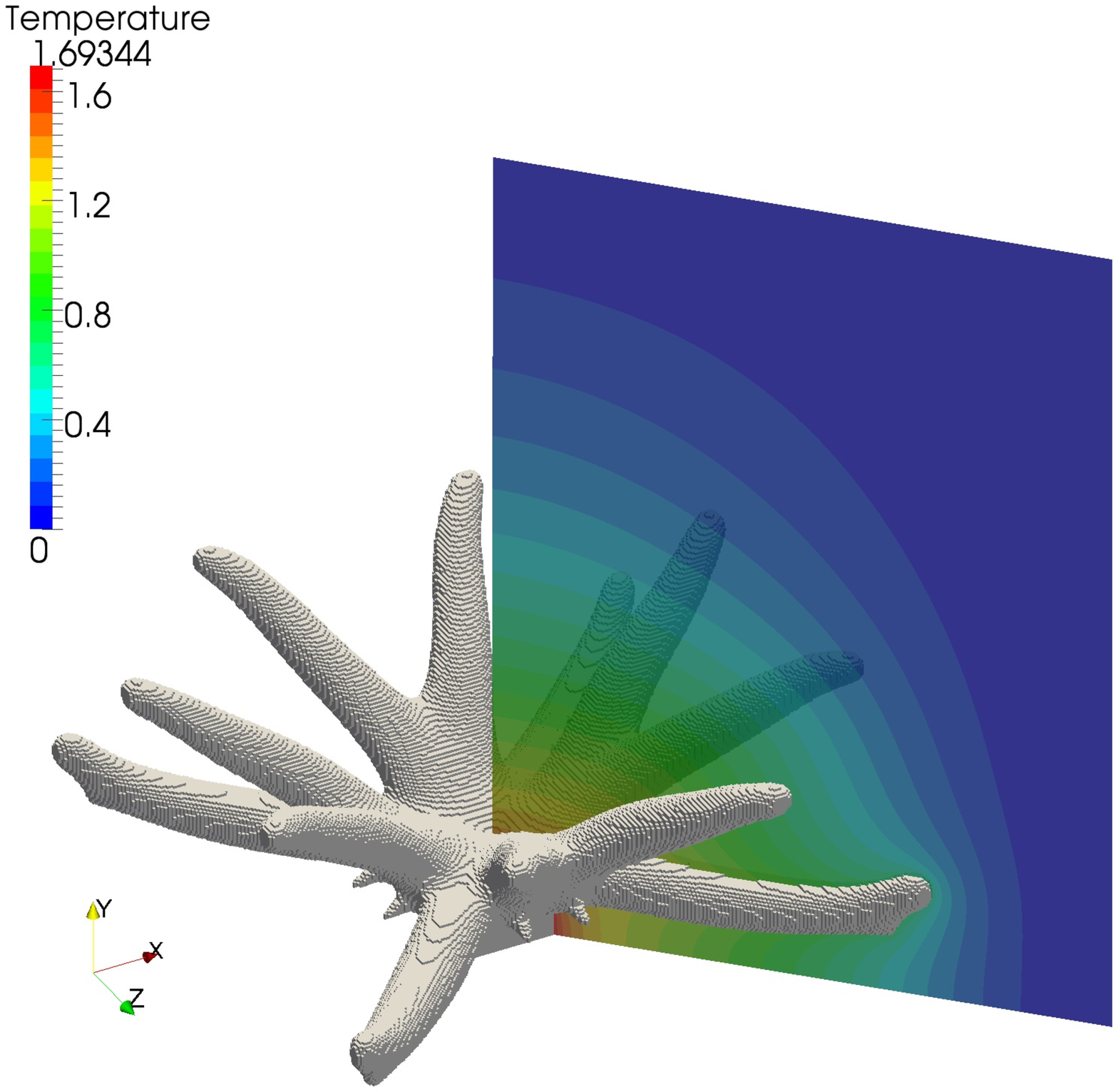}
\label{fig:optDes_N320_tempSlice-a}}
\subfloat[$Gr=10^{4}$]{\includegraphics[width=0.45\textwidth]{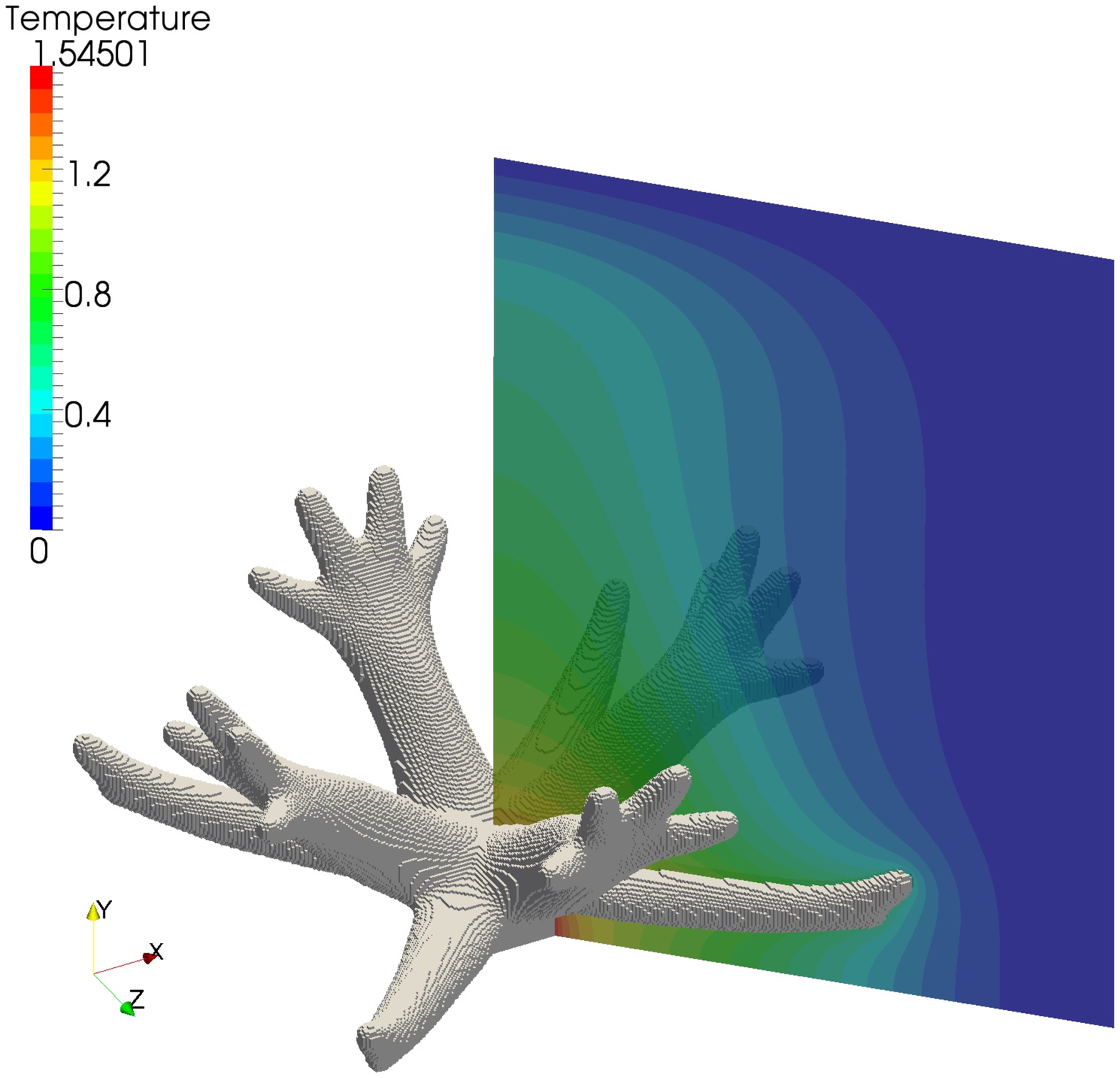}
\label{fig:optDes_N320_tempSlice-b}}\\
\subfloat[$Gr=10^{5}$]{\includegraphics[width=0.45\textwidth]{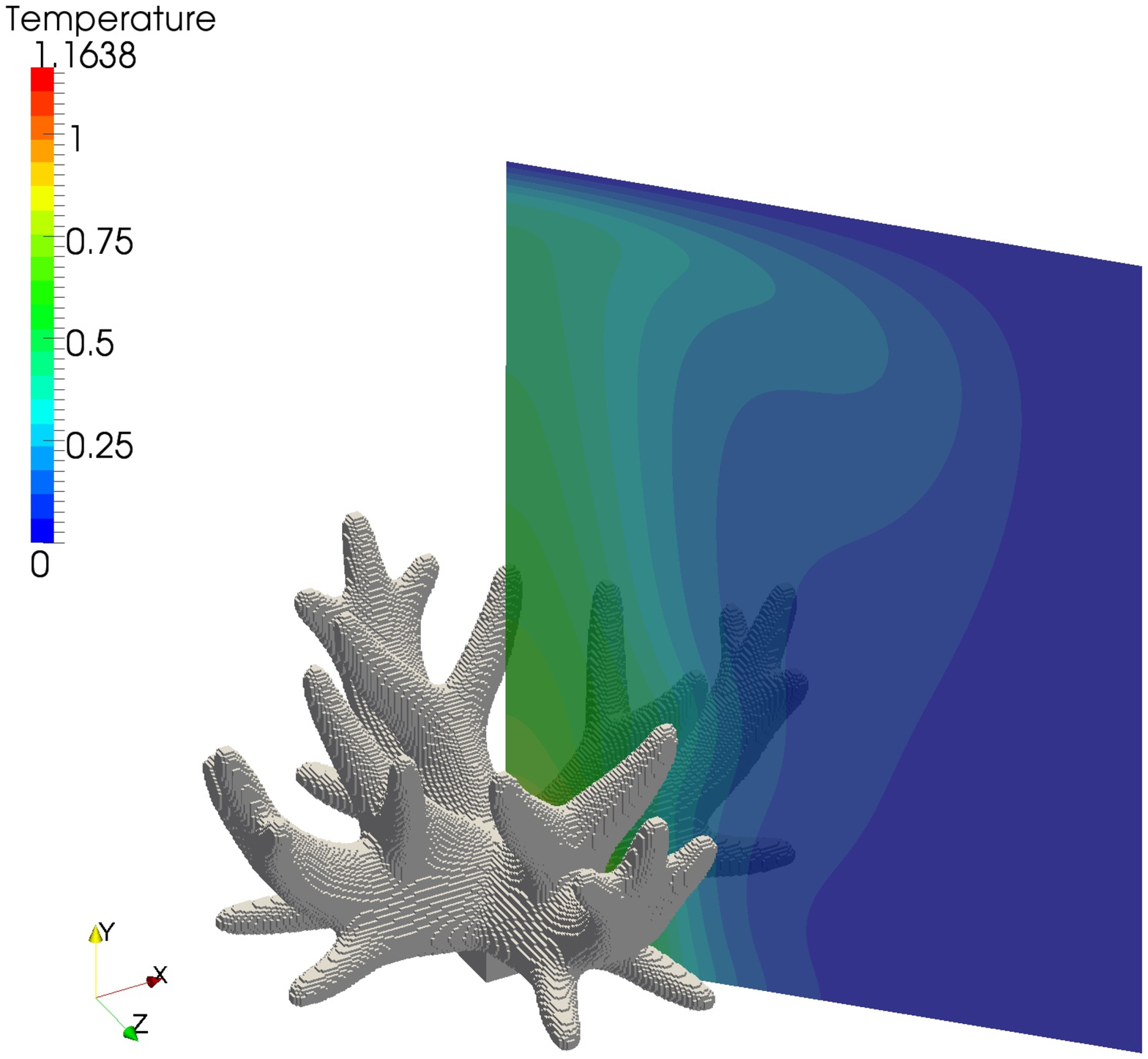}
\label{fig:optDes_N320_tempSlice-c}}
\subfloat[$Gr=10^{6}$]{\includegraphics[width=0.45\textwidth]{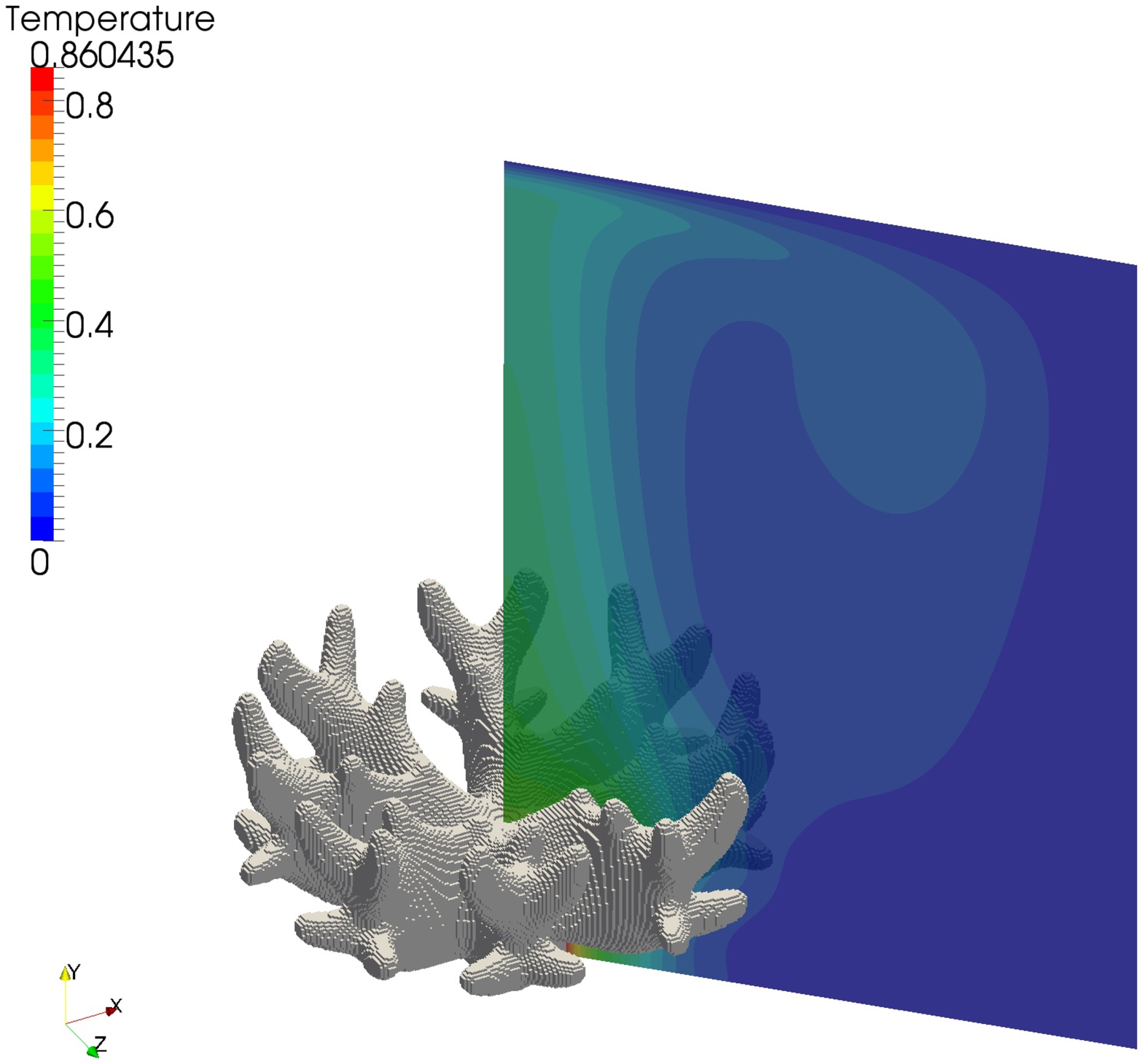}
\label{fig:optDes_N320_tempSlice-d}}
\caption{Optimised designs for varying $Gr$-number at a mesh resolution of $160\times320\times160$.}\label{fig:optDes_N320_tempSlice}
\end{figure*}
Figure \ref{fig:optDes_N320_tempSlice} shows the optimised designs for varying $Gr$-number with superimposed slices of the corresponding temperature fields. Due to the use of density filtering, the interface between solid and fluid regions for the optimized designs are not exactly described but consists of a thin layer of intermediate design field values. The optimised design distributions are shown thresholded at $\gamma=0.05$, which is the approximate location of the computational interface. 
The general characteristics of all the designs are similar, i.e. all designs are ``thermal trees'' with conductive branches moving the heat away from the heat source. However, it can clearly be seen that the design changes considerably with increasing convection-dominance (increasing $Gr$). For increasing $Gr$-number the conductive branches contract, resulting in a smaller spatial extent of the overall heat sink. This intuitively makes sense as the problem goes from one of conduction/diffusion at $Gr=10^{3}$ to convection at $Gr=10^{6}$. When diffusion dominates, the goal for the branches essentially becomes to conduct the heat directly towards the cold walls. As convection begins to matter, the fluid movement aids the transfer of heat away from the heat sink and the branches do not need to be as long. Instead, the design forms higher vertical interfaces in order to increase surface area perpendicular to the flow direction and thus better transfer of heat to the fluid.
At the same time the complexity of the designs increases as the importance of convection increases. This can be seen by studying the number of primary and secondary branches. Primary branches are defined as those connected to the heat source directly and secondary branches are those connected to primary branches.
\begin{table}
\centering
\begin{tabular}{cccc}
$Gr$  & Primary & Secondary & Surface area \\ \hline
$10^{3}$ & 12 &  0 & 0.887 \\
$10^{4}$ &  8 & 16 & 0.853 \\
$10^{5}$ &  8 & 28 & 0.834 \\
$10^{6}$ &  8 & 48 & 0.846 \\
\end{tabular}
\caption{The number of primary and secondary branches, as well as the surface area for the optimised designs of figure \ref{fig:optDes_N320_tempSlice}.} \label{tab:grVsBranches}
\end{table}
Table \ref{tab:grVsBranches} shows the number of primary and secondary branches of the optimised designs shown in figure \ref{fig:optDes_N320_tempSlice}. The number of primary branches is largest for the diffusion-dominated case, but more or less constant thereafter. However, it can be seen that the number of secondary branches significantly increases as the $Gr$-number is increased.
Table \ref{tab:grVsBranches} also shows that the total surface area\footnote{The surface area is computed using an isosurface at the selected threshold design field value.} is decreasing for the three lower $Gr$-numbers and then increases slightly.

It is interesting to note, that the trend of increasing complexity with increasing $Gr$-number is the reverse of what was observed for two-dimensional problems \cite{Alexandersen2014}. There, the complexity of the design decreased as the $Gr$-number increased. This difference is likely due to the fact that going into three-dimensions allows the fluid to move around and through the design making it more a question of ``topology'', in contrast to in two-dimensions where it is a question of surface shape. Physically, in two-dimensions additional branches disturb the flow and thus the heat transfer; in three-dimensions, vertical branches improve the heat transfer through an increased vertical surface area.
\begin{figure}
\centering
\subfloat[$Gr=10^{3}$]{\includegraphics[width=0.45\textwidth]{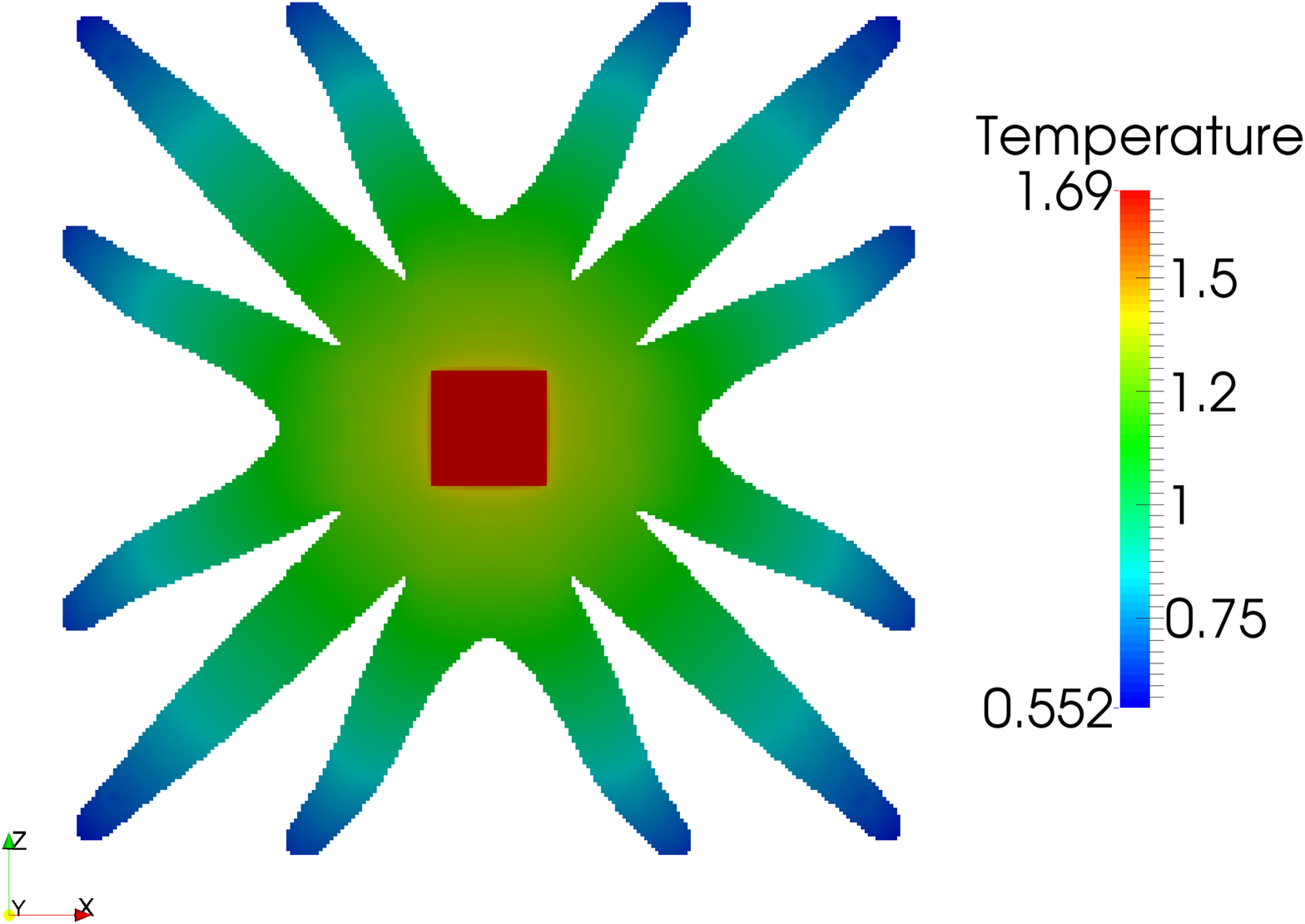}\label{fig:optDes_N320_bottomView-a}}\\
\subfloat[$Gr=10^{6}$]{\includegraphics[width=0.45\textwidth]{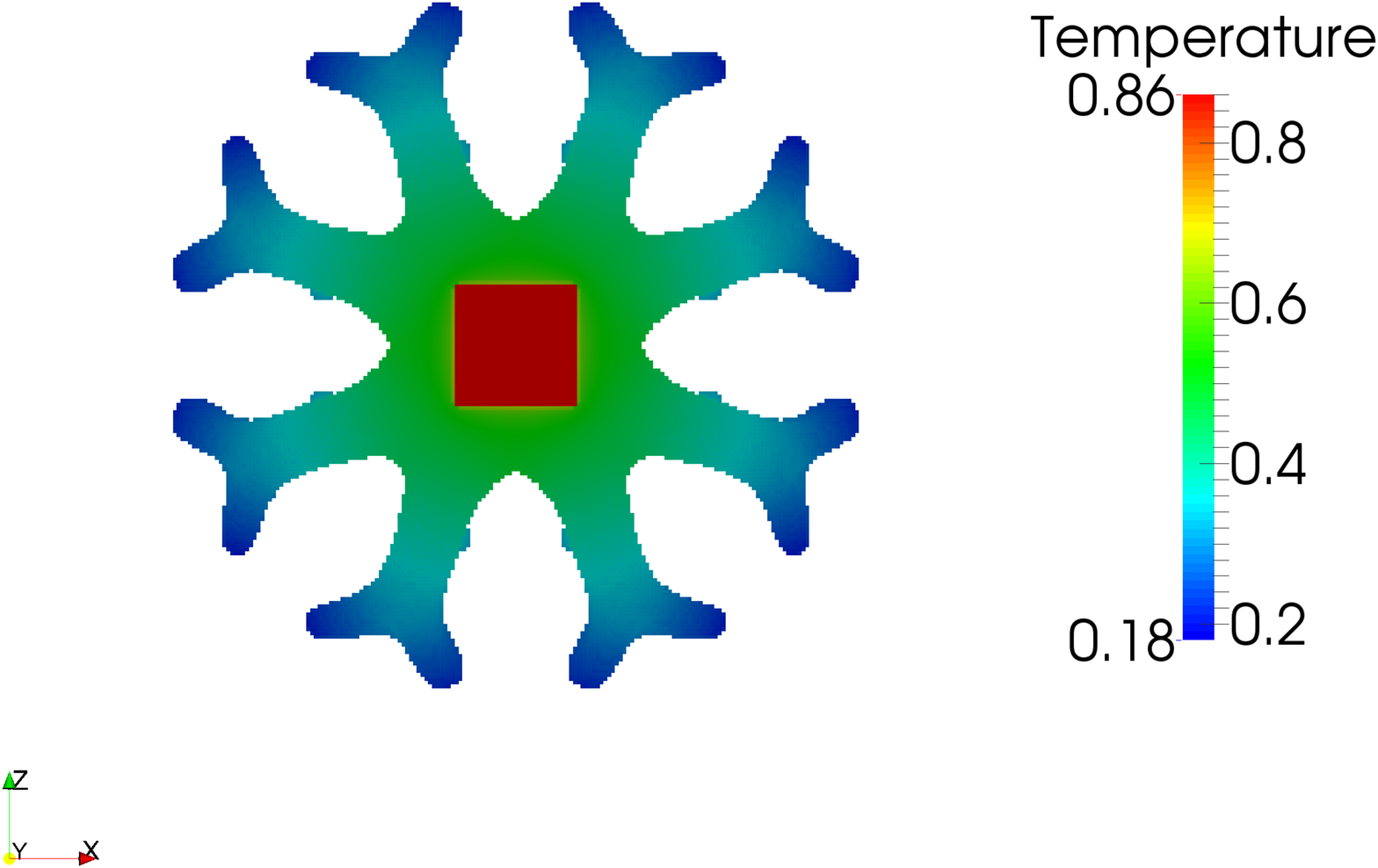}\label{fig:optDes_N320_bottomView-b}}
\caption{Temperature distribution in optimised designs for $Gr=10^{3}$ and $Gr=10^{6}$ at a mesh resolution of $160\times320\times160$ - view from below.} \label{fig:optDes_N320_bottomView}
\end{figure}
Figure \ref{fig:optDes_N320_bottomView} shows the optimised designs for $Gr=10^{3}$ and $Gr=10^{6}$ from below. The radial extent of the designs are emphasised from these views. It is also seen that the branches for $Gr=10^{6}$ are positioned to ensure that the structure is open from below, with the branches forming vertical walls as discussed above.

\begin{table}
\centering
\begin{tabular}{ccccc}
 & \multicolumn{4}{c}{Optimisation $Gr$} \\
Analysis $Gr$ & $10^{3}$ & $10^{4}$ & $10^{5}$ & $10^{6}$ \\ \hline
$10^{3}$ & \blue{2.064962} & 2.066856 & 2.241125 & 2.362470 \\
$10^{4}$ & 1.933460 & \blue{1.880946} & 1.994538 & 2.111726 \\
$10^{5}$ & 1.488278 & 1.450276 & \blue{1.404511} & 1.439089 \\
$10^{6}$ & 1.134963 & 1.121979 & 1.061452 & \blue{1.025340} \\
\end{tabular}
\caption{Crosscheck objective function values for the designs shown in figure \ref{fig:optDes_N320_tempSlice}.} \label{tab:crosscheck}
\end{table}
Table \ref{tab:crosscheck} shows a crosscheck of the objective functions for the optimised designs. It can be seen that all designs optimised for certain flow conditions outperform the other designs for the specified $Gr$-number.

The optimisations was run for 500 design iterations for each optimised design.
\begin{table}
\centering
\begin{tabular}{cccc}
$Gr$  & Time & Non-linear:  & Linear:  \\
      &      &  avg. (max)  &  contin - avg. (max) \\ \hline
$10^{3}$ &  9:56 & 1.9 (2) & 7.6,8.4,7.8,8.1,18.4 - 10.1 (25) \\
$10^{4}$ & 10:25 & 2.0 (3) & 8.3,8.6,8.3,8.6,22.7 - 11.3 (29) \\
$10^{5}$ & 10:28 & 2.1 (10) & 8.4,8.6,8.7,8.2,15.7 - 9.9 (34) \\
$10^{6}$ & 10:35 & 2.1 (7) & 7.3,7.4,7.5,8.0,8.4 - 7.7 (14) \\
\end{tabular}
\caption{Computational time, average non-linear iterations and linear iterations for the optimised designs of figure \ref{fig:optDes_N320_tempSlice}.} \label{tab:grVsPerformance}
\end{table}
Table \ref{tab:grVsPerformance} shows the computational time, average non-linear iterations and linear iterations for the optimised designs of figure \ref{fig:optDes_N320_tempSlice} using 1280 cores. It can be seen that the computational time only weakly increases as the $Gr$-number is increased and remains close to 1.2 minutes per design iteration on average. It is interesting to note, that $Gr=10^{6}$ seems easier to solve than the others as it exhibits lower average number of linear iterations than all others, as well as a lower maximum number of non-linear iterations than $Gr=10^{5}$. Furthermore, it can be seen that due to the Newton method starting from a good initial vector, only 2 non-linear iterations are needed for most of the design iterations independent of $Gr$-number.

\subsection{High resolution design} \label{sec:results_hugeDes}
The problem is now investigated with a computational mesh of $320\times640\times320$ elements yielding a total of $65,536,000$ elements and $330,246,405$ degrees of freedom for the quarter domain. The design domain consists of $27,648,000$ elements and the filter radius is set to $2.5$ times the element size, i.e. in absolute measures half the size of before.
The optimisation is run for 1000 design iterations and the computational time was 107 and 108 hours, respectively, for $Gr=10^{3}$ and  $Gr=10^{6}$, using 2560 cores. This yields an average of 6.4 and 6.5 minutes per design iteration, respectively.

\begin{figure*}
\centering
\subfloat[Iteration 200]{\includegraphics[width=0.45\textwidth]{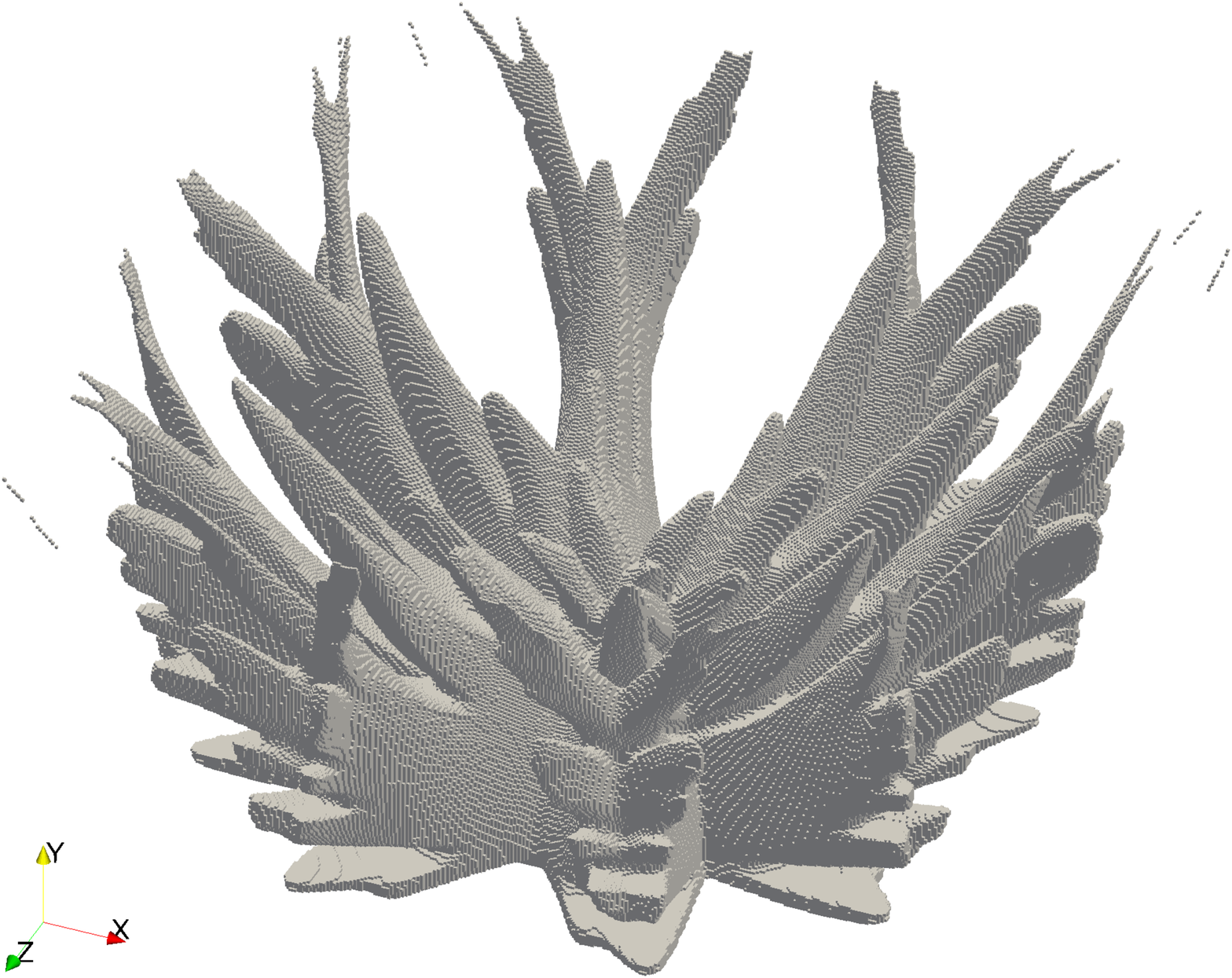}
\label{fig:optDes_N640_Gr1e6-a}}
\subfloat[Iteration 400]{\includegraphics[width=0.45\textwidth]{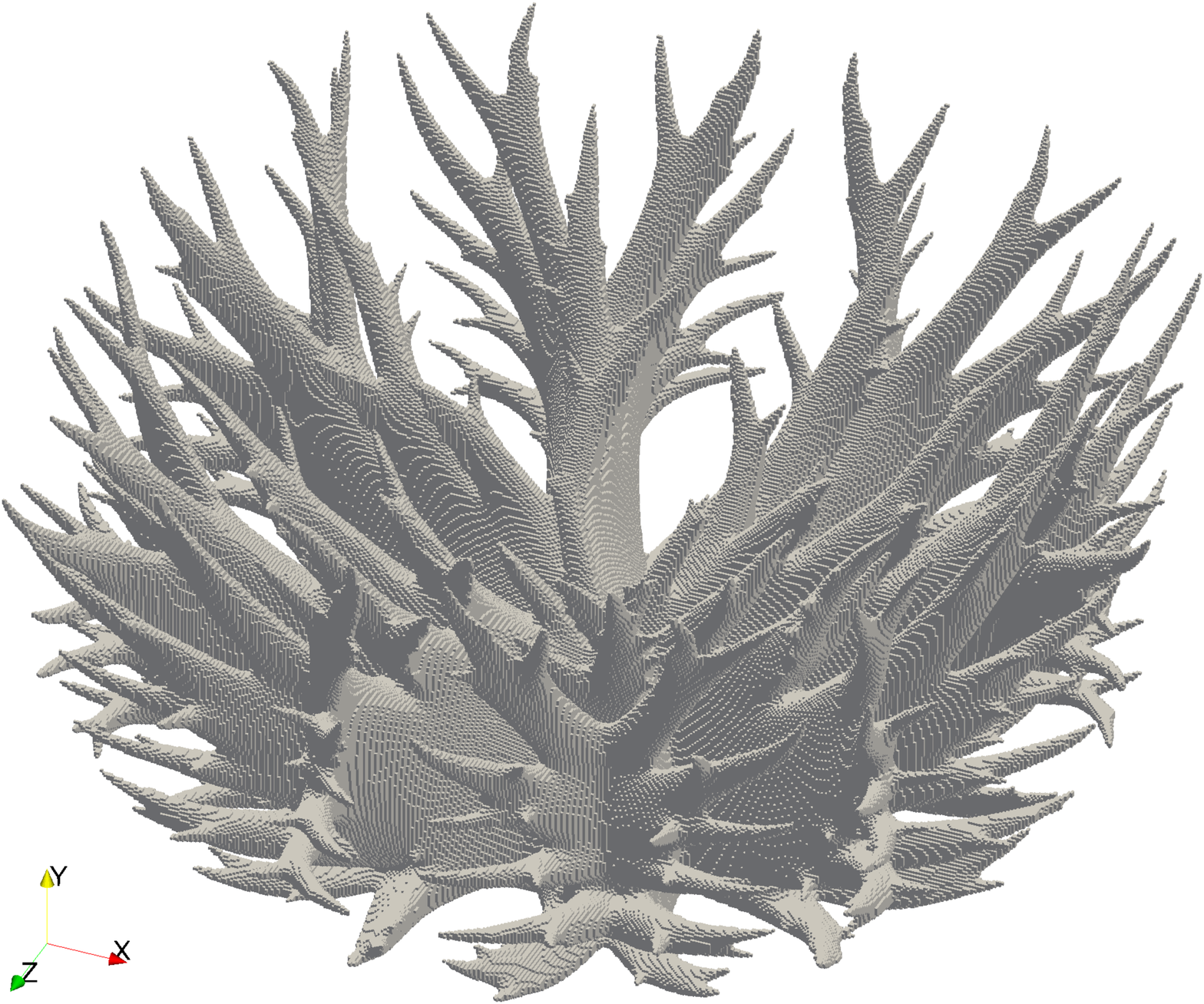}
\label{fig:optDes_N640_Gr1e6-b}} \\
\subfloat[Iteration 600]{\includegraphics[width=0.45\textwidth]{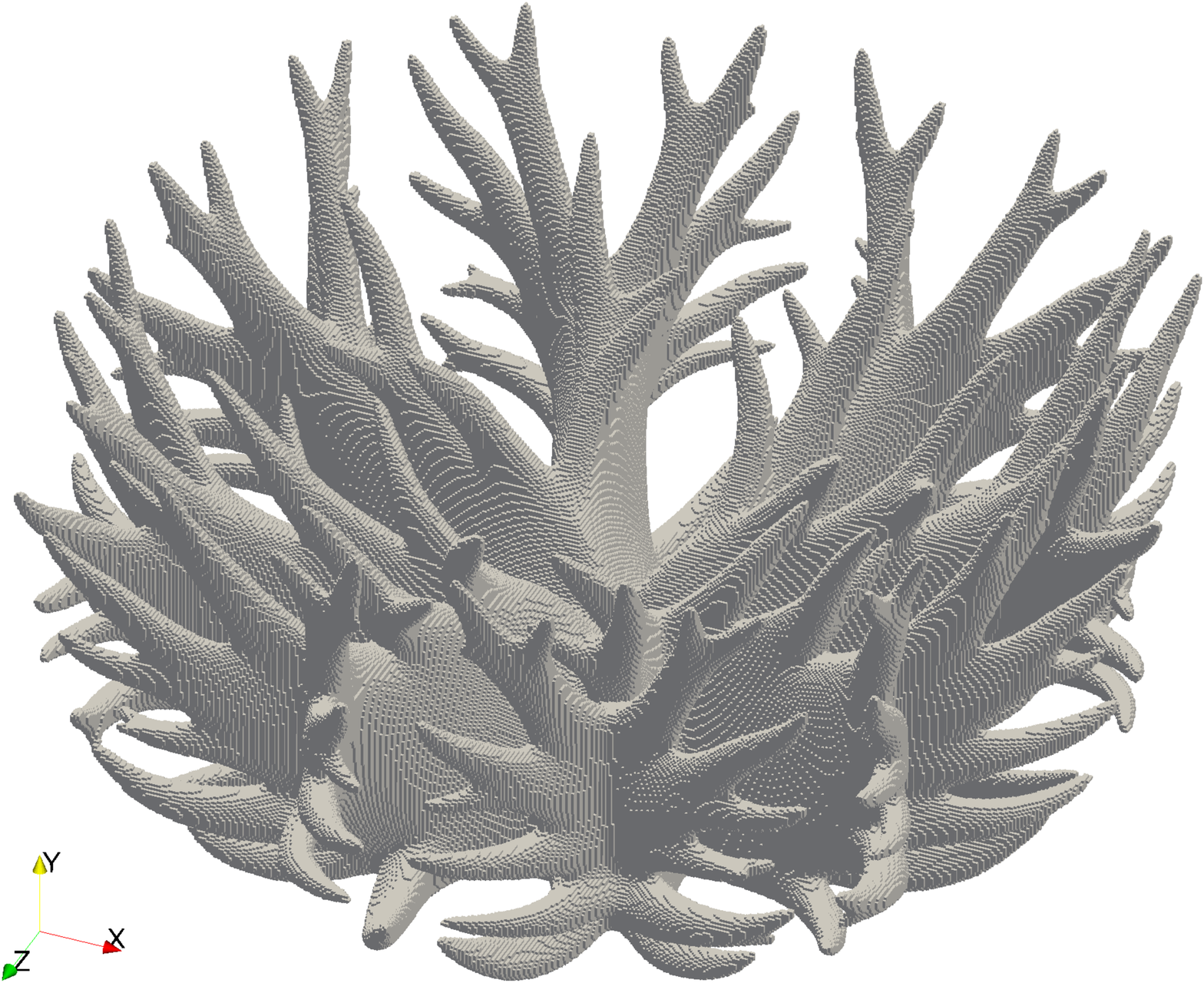}
\label{fig:optDes_N640_Gr1e6-c}}
\subfloat[Iteration 1000]{\includegraphics[width=0.45\textwidth]{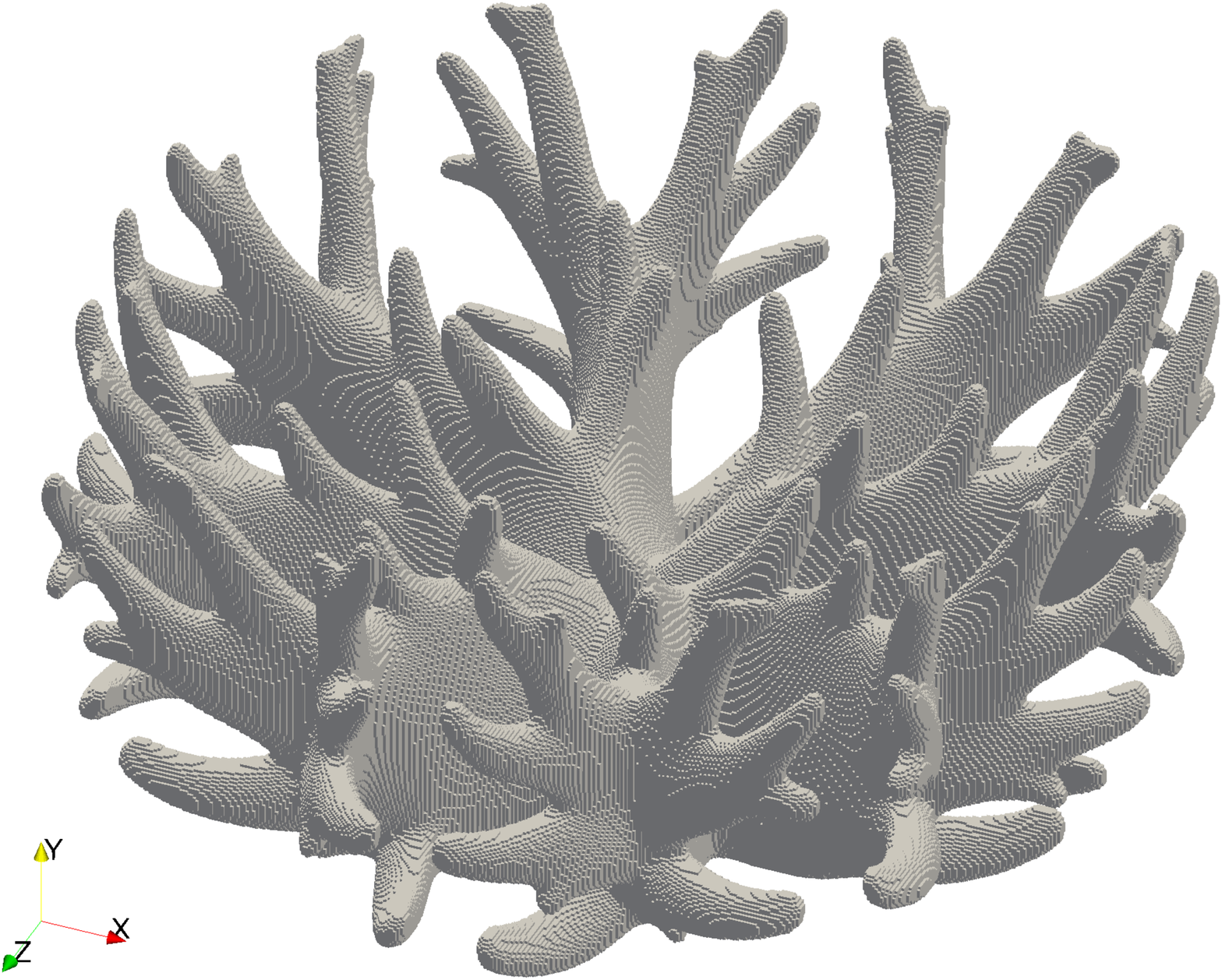}
\label{fig:optDes_N640_Gr1e6-d}}
\caption{Optimised designs for $Gr=10^{6}$ at a mesh resolution of $320\times640\times320$. Please note that the freely hanging material in (a) is due to only elements below the threshold, $\gamma=0.05$, are shown - the design is connected by intermediate design field values.}\label{fig:optDes_N640_Gr1e6}
\end{figure*}
Figure \ref{fig:optDes_N640_Gr1e6} shows the optimised design for $Gr=10^{6}$ with the fine mesh resolution and small length scale at various steps of the continuation strategy. The complexity of the design can be seen to be significantly higher than for the design with a larger length scale, figure \ref{fig:optDes_N320_tempSlice-d}. The subfigures are the final iterations of the 1st, 2nd, 3rd and 5th (final) continuation steps. It can be seen that the complexity of the design decreases during the optimisation process once the overall topology has been settled (iteration 400 and onwards). This is due to the harder and harder penalisation of intermediate design field values, with respect to conductivity, which are present at the interface between solid and fluid. Therefore, smaller features are progressively removed as the surface area is more heavily penalised. The reason for going to such high penalisation of the conductivity is to ensure the approximate collocation of the fluid and thermal boundaries. However, if one starts directly with these physically-relevant parameters, particularly poor local minima have been observed.

\begin{figure*}
\centering
\subfloat[$Gr=10^{3}$]{\includegraphics[width=0.45\textwidth]{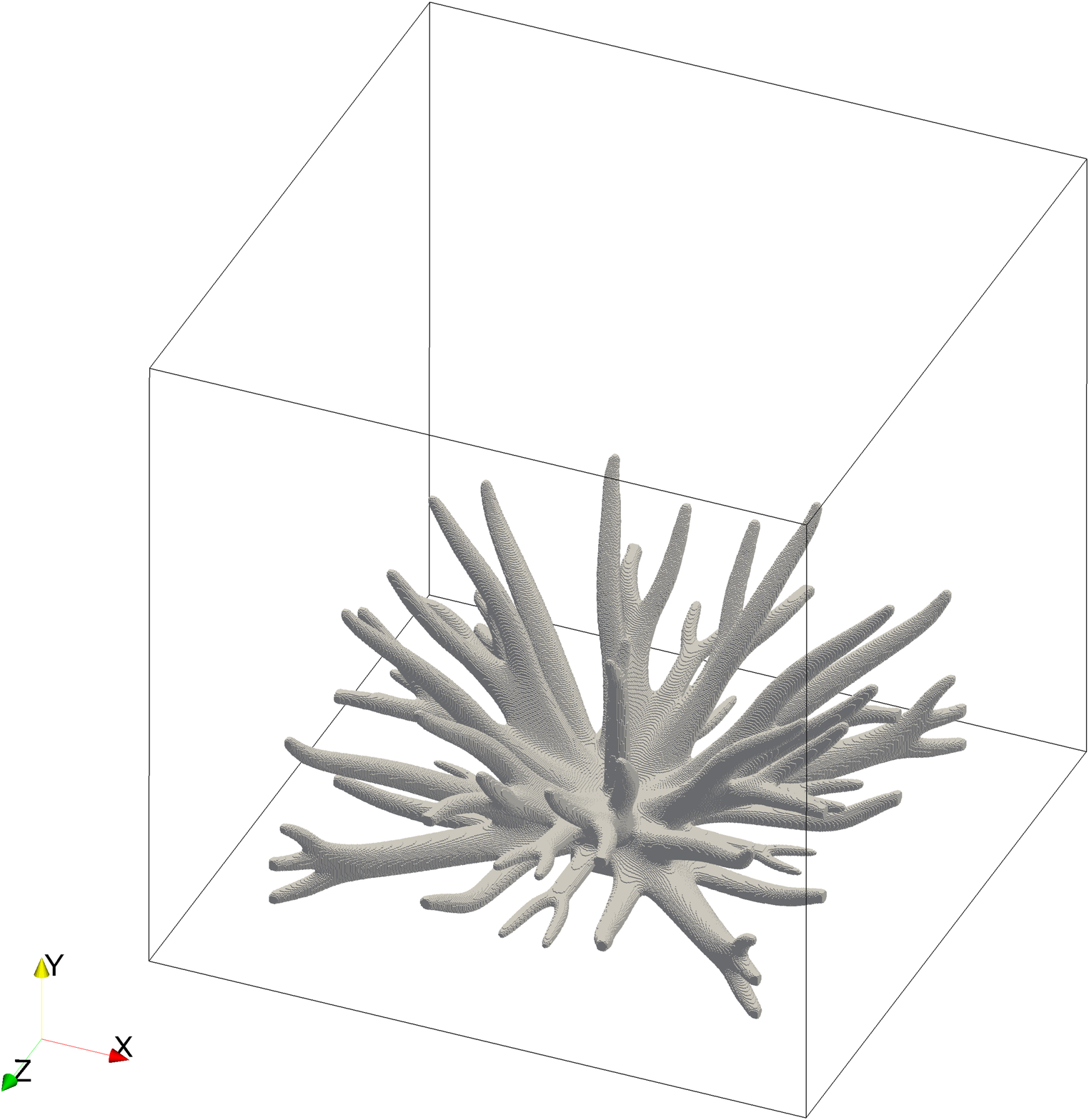} \label{fig:optDes_N640-a}}
\subfloat[$Gr=10^{6}$]{\includegraphics[width=0.45\textwidth]{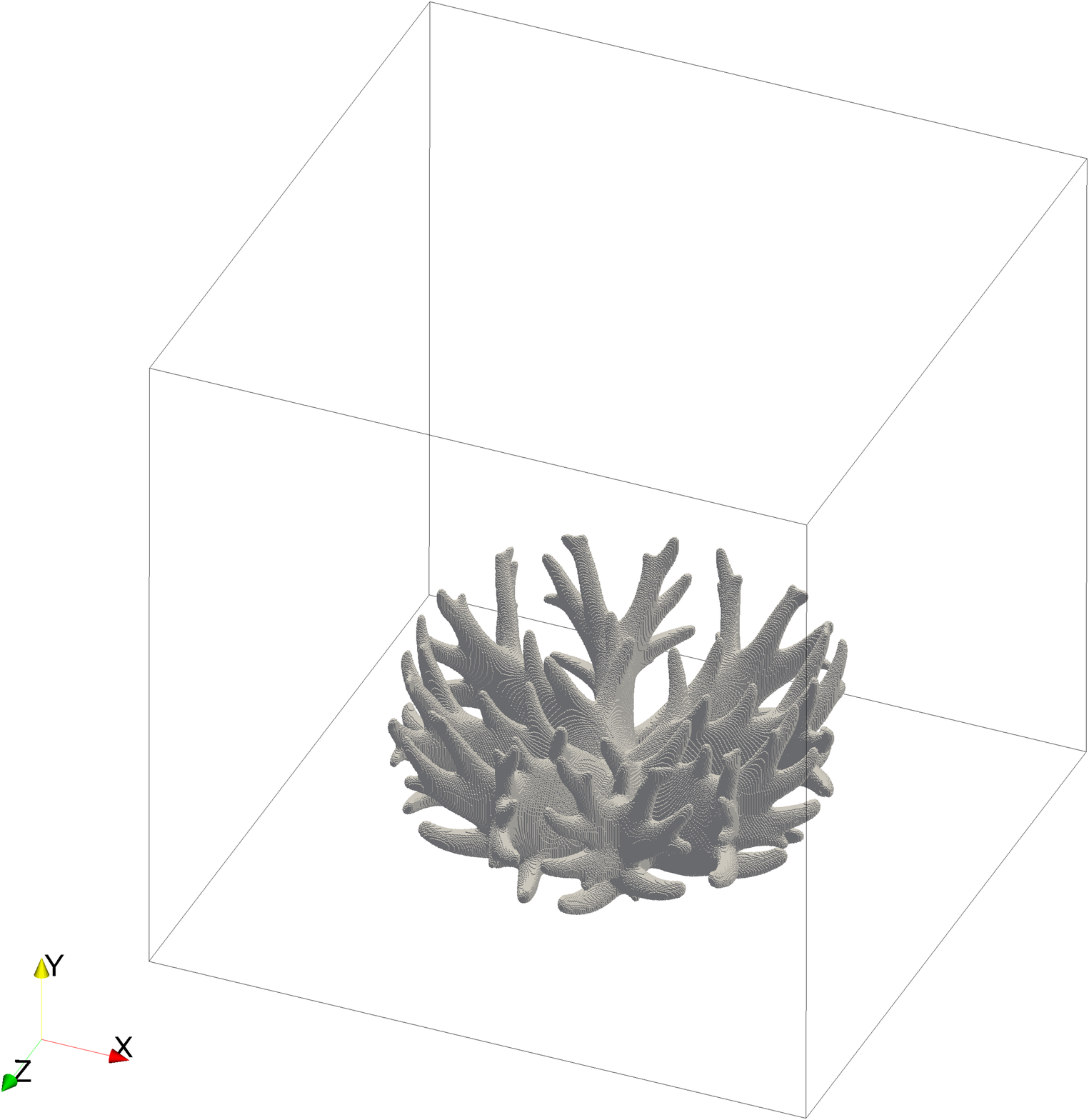} \label{fig:optDes_N640-b}}
\caption{Optimised designs for $Gr=10^{3}$ and $Gr=10^{6}$ at a mesh resolution of $320\times640\times320$. The outline of the outer walls are shown in black.} \label{fig:optDes_N640}
\end{figure*}
Figure \ref{fig:optDes_N640} shows the final optimised designs for $Gr=10^{3}$ and $Gr=10^{6}$ with the fine mesh resolution and small length scale. The complexities of both designs can be seen to be significantly higher than the previous, figures \ref{fig:optDes_N320_tempSlice-a} and \ref{fig:optDes_N320_tempSlice-d}, due to the smaller length scale. The increase in complexity with increasing Grashof number is not as apparent at this resolution and length scale, however, it is still argued that the number of primary and secondary members follows the trend from the previous study, section \ref{sec:results_varGr}. Also, the shift from long conducting branches to shorter members, with significant vertical surface area, is still obvious.

\section{Discussion and conclusion} \label{sec:conclusions}
In this paper we have applied topology optimisation to the design of three-dimensional heat sinks using a fully coupled non-linear thermofluidic model. In contrast to previous works, that considered simplified convection models, the presented methodology is able to recover interesting physical effects and insights, and avoids problems with the formation of non-physical internal cavities, length-scale effects and artificial convection assumptions. The implementation of the code in a PETSc framework suitable for large scale parallel computations allows for running examples with more than 300 million degrees of freedom and almost 30 million design variables on regular grids.

The example considered in the paper was primarily of academic nature. Nevertheless, some interesting insight is obtained, showing that optimised structures go from exhibiting simple branches that conduct heat towards the cold outer boundaries for diffusion-dominated problems, towards complex and compact multi-branched structures that maximize the convection heat transfer for higher Grashof numbers.

Current and future work includes applications to real life problems (see preliminary work in \citep{Alexandersen2015a}), irregular meshes, multiple orientations, as well as the extension to transient problems and detailed investigations into the modelling accuracy of the boundary layer.

\section{Acknowledgements} \label{sec:acknow}
This work was funded by Villum Fonden through the NextTop project, as
well as by Innovation Fund Denmark through the HyperCool project. The first author was also partially funded by
the EU FP7-MC-IAPP programme LaScISO.

\appendix

\section{Computational versus actual Grashof number} \label{app:grashofNumber}
The Grashof numbers used above are all based on an \textit{a priori} defined reference temperature difference. However, the actual temperature difference observed between the heat source and the walls are not known \textit{a priori} due to the fact that the problem only has a single known temperature (Dirichlet boundary condition) and a volumetric heat source. This is why the maximum temperature observed for the designs, see figure \ref{fig:optDes_N320_tempSlice}, is not equal to 1.
Therefore, after the optimisation, one can define an \textit{a posteriori} Grashof number based on the actual computed temperature difference. The \textit{a priori} version can be termed the computational Grashof number, the one that goes into the dimensionless governing equations; and the \textit{a posteriori} version can be called the actual, or optimised, Grashof number for the given optimised design. The actual Grashof number can be useful for experimental studies, as well as for future comparisons.
\begin{table}
\centering
\begin{tabular}{cccc}
Computational $Gr$  	& Mesh 						& Figure  							& Actual $Gr$  \\ \hline
$10^{3}$ 			& $160\times320\times160$ 	& \ref{fig:optDes_N320_tempSlice-a} 	& $1.69\times10^{3}$ \\
$10^{4}$ 			& $160\times320\times160$ 	& \ref{fig:optDes_N320_tempSlice-b}	& $1.55\times10^{4}$ \\
$10^{5}$ 			& $160\times320\times160$ 	& \ref{fig:optDes_N320_tempSlice-c}	& $1.16\times10^{5}$ \\
$10^{6}$ 			& $160\times320\times160$ 	& \ref{fig:optDes_N320_tempSlice-d}	& $8.60\times10^{5}$ \\
$10^{3}$ 			& $320\times640\times320$ 	& \ref{fig:optDes_N640-a} 		& $1.42\times10^{3}$ \\
$10^{6}$ 			& $320\times640\times320$ 	& \ref{fig:optDes_N640-b}		& $7.16\times10^{5}$
\end{tabular}
\caption{Computational and actual $Gr$-numbers for the designs presented in this paper.} \label{tab:compVsActGr}
\end{table}
The actual Grashof numbers for the designs shown in this paper are listed in Table \ref{tab:compVsActGr}. The values for the fine meshes are lower due to the lower thermal compliance, equivalent to the temperature of the heat source, allowed by the smaller design length scale.

\section{Stabilisation parameters} \label{app:stabilisation}
The stabilised weak form of the equations are given in \cite{Alexandersen2014}. The equations are stabilised using the pressure-stabilising Petrov-Galerkin (PSPG) \cite{Hughes1986a,Tezduyar1992a} and streamline-upwind Petrov-Galerkin (SUPG) methods \cite{Brooks1982}, for more information please see \cite{Alexandersen2014}.

The stabilisation parameters for SUPG and PSPG are defined as one and the same using the following approximate min-function:
\begin{equation} \label{eq:stabFacs_globalDefinition}
\tau_{SU} = \tau_{PS} = \tau = \left( \frac{1}{{\tau_{1}}^{2}} + \frac{1}{{\tau_{2}}^{2}} + \frac{1}{{\tau_{3}}^{2}} \right)^{-\frac{1}{2}}
\end{equation}
The limit factors are given by:
\begin{subequations} \label{eq:stabFacs_SimpleDefinition}
\begin{align}
\tau_{1} &= \frac{4 h_{e}}{\norm{\vecr{u}_{e}}{2} } \\
\tau_{2} &= \frac{{h_{e}}^{2}}{4 Pr} \\
\tau_{3} &= \frac{1}{\alpha_{e}}
\end{align}
\end{subequations}
where $h_{e}$ is a characteristic element size (for cubes the element edge length) and $\vecr{u}_{e}$ is the element vector of velocity degrees of freedom. The first limit factor, $\tau_{1}$, has been simplified based on evaluation at element centroids under the assumption of a single Gauss-point, yielding a constant stabilisation factor within each element.

In order to define a consistent Jacobian matrix, and thus a consistent adjoint problem, the derivatives of the stabilisation factors are needed with respect to the velocity field.
This can be found to be:
\begin{equation}
\dpart{\tau}{\vecr{u}_{e}} = -\tau \left( 1 + \left(\frac{{\tau_{1}}}{{\tau_{2}}}\right)^{2} + \left(\frac{{\tau_{1}}}{{\tau_{3}}}\right)^{2} \right)^{-1} \left( \T{\vecr{u}_{e}}\vecr{u}_{e} \right)^{-1} \T{\vecr{u}_{e}}
\end{equation}

Furthermore, to define consistent design sensitivities, the derivatives of the stabilisation factors with respect to the design field is needed. This can be found to be:
\begin{equation}
\dpart{\tau}{\gamma_{e}} = - \frac{\tau}{\tau_{3}} \left( \left(\frac{{\tau_{3}}}{{\tau_{1}}}\right)^{2} + \left(\frac{{\tau_{3}}}{{\tau_{2}}}\right)^{2} + 1 \right)^{-1} \dpart{\alpha_{e}}{\gamma_{e}}
\end{equation}

Including the derivatives in the definition of consistent adjoint sensitivities has been observed to provide a one to two order of magnitude improvement in accuracy of sensitivities with respect to a finite difference approximation. Significant differences have not been observed in optimisation behaviour or in final designs. However, this cannot be guaranteed in general and it is therefore best to ensure consistency, as also highlighted by the similar issue of frozen turbulence \cite{Zymaris2009}.

Similar definitions and derivations are carried out for the thermal SUPG stabilisation.


\bibliographystyle{elsarticle-num}
\bibliography{bib}

\end{document}